\documentclass[12pt]{article}
\usepackage{a4wide,epsfig,amsmath,amssymb,cite,scalefnt,rccol,fltpoint,thophys}
\begin{document}

\title{
\begin{flushright}
 \normalsize{TTP09-20}\\
 \normalsize{SFB/CPP-09-57}\\
\end{flushright}
\vspace{1.5cm}
  Supersymmetric Corrections to the Threshold Production of Top Quark Pairs}

\author{\small Y. Kiyo, M. Steinhauser, N. Zerf\\
  {\small\it Institut f{\"u}r Theoretische Teilchenphysik,
    Universit{\"a}t Karlsruhe,}\\
  {\small\it Karlsruhe Institute of Technology (KIT),}\\
  {\small\it 76128 Karlsruhe, Germany}\\
}

\date{}

\maketitle

\thispagestyle{empty}

\begin{abstract}
In this paper we investigate supersymmetric effects
to the threshold production
cross section of top quark pairs in electron positron annihilation.
In particular, we consider the complete one-loop corrections from
the strong and weak sector of the Minimal Supersymmetric Standard Model.
\medskip


\end{abstract}


\section{Introduction}

One of the main goals of a future electron positron collider is the
precise measurement of the top quark production cross section in the
threshold region. The comparison to the theoretical prediction allows for
a precise extraction of the top quark mass, its width, the strong
coupling and --- in case the Higgs boson is not too heavy --- the top
quark Yukawa coupling.

The next-to-next-to-leading order (NNLO) QCD corrections to the total cross
section $\sigma(e^+e^-\to t\bar{t})$
has been completed several years ago~\cite{Hoang:2000yr}. One observes large
perturbative corrections from the second 
order terms which make a precise prediction difficult. In the recent years a
big effort has 
been undertaken to complete the third-order corrections to
$\sigma(e^+e^-\to t\bar{t})$ 
\cite{Kniehl:2002br,Penin:2005eu,Beneke:2005hg,Marquard:2006qi,Eiras:2006xm,Beneke:2007gj,Beneke:2007pj,Beneke:2008cr,Kiyo:2008mh,Marquard:2009bj,Smirnov:2008pn} .
First numerical estimates \cite{Beneke:2008ec} indicate that the convergence of the
perturbation theory is improved after the inclusion of the NNNLO
terms. In addition to the third-order corrections also the
resummation of the NNLL terms is
studied~\cite{Hoang:2000ib,Hoang:2001mm, Pineda:2006ri}.

In order to profit from precise experimental measurements it is
desired to reach an uncertainty below approximately 3\% from the theory
side~\cite{Martinez:2002st}.
Radiative corrections of this order can easily be reached by effects
from theories beyond the Standard Model (SM). In this paper we consider the
effect of supersymmetric corrections within the
Minimal Supersymmetric Standard Model (MSSM). Furthermore we
confirm the results from
Refs.~\cite{Grzadkowski:1986pm,Guth:1991ab,Denner:1991tb} and
\cite{Hoang:2006pd} obtained in the 
framework of the SM and two-Higgs-doublet model (THDM) of type II,
respectively. 

It is convenient to perform the calculation of the production cross
section in the framework of an effective theory where the
produced top quarks are described by a non-relativistic two-particle
Green's function. All effects connected to energy scales above $\mu
\approx m_W$ are contained in coefficient functions which
represent the new couplings in the effective Lagrangian.
Since the masses of the supersymmetric particles are above the electroweak scale
they only influence the matching coefficients of the effective operators. For
the top quark
production we have to consider the vector current in the full and effective
theory which constitutes a building block for all threshold phenomena
involving the coupling of the initial electron and positron via photon, $Z$
boson or box diagrams to heavy quarks. 

The remainder of the paper is organized as follows: In the next
Section we provide the formulae which are necessary for the evaluation of the
threshold cross section. Afterwards we discuss in Sections~\ref{sec::sqcd}
and~\ref{sec::ew} the numerical effects from the strong and weak sector of the
MSSM and present in Section 5 our conclusions.


\section{Framework}

Non-relativistic QCD (NRQCD) allows for a consistent separation of
the hard corrections connected to energy scales of the order
of the weak gauge bosons or higher from the soft scales which
are involved in the top anti-top boundstate.
Within NRQCD we can normalize the production cross section to
$\sigma(e^+e^-\rightarrow \mu^+\mu^-)=(4\pi\alpha^2)/(3s)$
and denote the ratio by $R$
\begin{eqnarray}
  R\left(e_{L}^+ e_{R}^-  \rightarrow t\bar{t} X\right)
  &=&
  \frac{8\pi}{s}\,
  {\rm Im}
  \big[\, \left(h_{R,V}\right)^{\,2} H_{V}
  +\left(h_{R,A}\right)^{\,2} H_{A} \big]\,,
  \label{eq:XSection}
\end{eqnarray}
where $s$ is the square of the center-of-mass energy.
In Eq.~(\ref{eq:XSection}) left-handed positrons
and right-handed electrons are considered;
for $e^+_R e^-_L$ in the initial state a similar expression
is obtained by replacing R by L in Eq.~(\ref{eq:XSection}).
Note that the initial states $e^+_R e^-_R$ and $e^+_Le^-_L$ are
suppressed by a factor $(m_e/M_W)^2\sim 10^{-10}$ and are thus negligible.
$h_{R,V}$ and $h_{R,A}$ are so-called helicity amplitudes which absorb the matching
coefficients representing the coupling of the effective operators.
They take care of the hard part of the reaction.
The first subscript of $h$ refers to helicity
of the electron, and the second one to the
vector ($J_{V}^\mu=\bar{\psi}\gamma^\mu\psi$) or
axial-vector coupling ($J_{A}^\mu=\bar{\psi}\gamma^\mu\gamma_5\psi$)
of the gauge bosons to the top quark current. 
In this paper we evaluate corrections to $h_{R,V}$ and $h_{L,V}$.

The bound-state dynamics is contained in the so-called hadronic part formed by
current-current correlators within NRQCD. They are
denoted by $H_V$ and $H_A$ in Eq.~(\ref{eq:XSection}) and will not
be considered further in this paper. At threshold the contribution from the
axial-vector current is 
suppressed by two powers of top quark velocity thus we only consider the 
vector current $J_{V}^\mu$ in this work. 
Its counterpart in the effective theory reads
$j_V^{\,i} = \psi^\dag\,\sigma^i\chi$.

It is convenient to separate the photon and $Z$ contribution in $h_{I,V}^{\rm tree}$ and
write
\begin{eqnarray}
  h_{I,V}^{\rm tree} &=& h_{I,V}^{\gamma,\rm tree} + h_{I,V}^{Z,\rm tree}\,.
\end{eqnarray}
Here the tree-level contributions are given by $(I=L/R)$
\begin{eqnarray}
  h_{I,V}^{\gamma,\rm tree} &=& Q_e Q_t\,,
  \nonumber\\
  h_{I,V}^{Z,\rm tree} &=& \frac{s \,\beta_I^{\,e}
    \,\beta_V^{\,t}}{s-M_Z^2}\,,
  \nonumber\\
  \beta_V^{\,t}&=&\frac{\beta_R^{\,t}+\beta_L^{\,t}}{2}\,,
  \nonumber \\
  \beta_{I}^f &=&\frac{(T_3)_{\,f_I}-s_w^2 Q_f}{s_w c_w}\,,
\label{eq:hamp}
\end{eqnarray}
where the $\beta_{I}^{f}$ is the coupling of a fermion ($f=e,t$)
to the $Z$ boson, $s_w$  is the sine of the weak mixing angle
($c_w^2=1-s_w^2=m_W^2/m_Z^2$),
and electric and iso-spin charges for top quark and electron are
given by
\begin{eqnarray}
  Q_e=-1, ~~ Q_t=2/3, ~~
  (T_3)_{\,t_L}=1/2, ~~(T_3)_{\,e_L}=-1/2,~~
  (T_3)_{\,f_R}\equiv 0.
  \label{eq:charges}
\end{eqnarray}
In the following the abbreviation $T_3^f\equiv (T_3)_{f_L}$ will be used.
Let us note that $h_{I,A}$ can be obtained by substituting $\beta_V^t$ by
$\beta_A^{\,t}=(\beta_R^t-\beta_L^t)/2$ in formula~(\ref{eq:hamp}).
The loop corrections are taken into account via
\begin{eqnarray}
  h_{I,V} &=& h_{I,V}^{\rm tree} + h_{I,V}^{\rm X}\,,
  \label{eq:corr2h}
\end{eqnarray}
where $X$ stands for QCD, SQCD (supersymmetric QCD), SM, THDM\footnote{In this
  paper we use the THDM type II where $u/d$-type quarks couple to different
  Higgs doublets $H_u/H_d$. Note that the Higgs sector of the MSSM corresponds
  to the Higgs sector of THDM type II.} or MSSM. The 
numerical effects are discussed for the quantity
\begin{eqnarray}
  \Delta^{\rm X} &=& \frac{\delta R^{\rm X}}{R^{\rm LO}}
  \,\,=\,\,
  \frac{2h_{L,V}^{\rm tree}\mbox{Re}\left(h_{L,V}^{\rm X}\right)
    + 2h_{R,V}^{\rm tree}\mbox{Re}\left(h_{R,V}^{\rm X}\right)}
  {\left(h_{L,V}^{\rm tree}\right)^2
    + \left(h_{R,V}^{\rm tree}\right)^2}
  \,,
\end{eqnarray}
where the sum over all helicity states of the incoming electron and positron
has been performed.
Let us note that in our case for the evaluation of $h_{I,V}$ one has to set $s=4m_t^2$.
Furthermore the external top quarks are on their mass shell.
In addition we are only interested in hard corrections resulting from the real
matching condition. Corrections to the cross section stemming from imaginary
part of the matching coefficient, which takes into account the finite lifetime
of the top quark, are discussed for SM in Ref. \cite{Hoang:2004tg}. 

For the generation of the Feynman diagrams
we use the {\tt Mathematica} program {\tt FeynArts}~\cite{Hahn:2000kx}.
The amplitudes are further processed with the help of the programs
{\tt FormCalc}~\cite{Hahn:1998yk} and {\tt FeynCalc}~\cite{Mertig:1990an}
which take the traces, map the occuring integrals to a standard basis
and reduce the tensor integrals to a minimal set of scalar integrals
usually denoted by $A_0$, $B_0$ and $C_0$. Since we have a quite
particular momentum configuration it is not possible to use the above
mentioned packages as black boxes but apply some modifications.
In fact, the choice $s=4m_t^2$ allows for a partial fractioning
in the denominators of the loop integrands appearing in
$t\overline{t}\gamma/Z$-vertex and box diagrams which effectively reduces the 
number of external legs by one. 
Consider, e.g., the integrand of a generic three-point function (omitting the
$i\epsilon$ prescription) 
\begin{eqnarray}
  \frac{1}{(p^2+2q_1p-M_1^2+m_t^2)(p^2-2q_2p-M_2^2+m_t^2)(p^2-M_3^2)}
  \,,
\end{eqnarray}
where $p$ is the integration momentum and $q_1^2=q_2^2=m_t^2$ are
the squared momenta of the top quarks. After
choosing $q_1=q_2=q/2$ and applying a partial fractioning one arrives at
\begin{eqnarray}
  \frac{2}{\frac{M_1^2+M_2^2}{2}-M_3^2-m_t^2}
  \left(\frac{1}{p^2-\frac{M_1^2+M_2^2}{2}+m_t^2}-\frac{1}{p^2-M_3^2}\right)
  \nonumber\\\qquad \times
  \left(\frac{1}{p^2+q\cdot p-M_1^2+m_t^2}+\frac{1}{p^2-q\cdot p-M_2^2+m_t^2}\right)
  \,.
\end{eqnarray}
As a consequence the result can be expressed in terms of only two-point
functions.
In a similar way one can express the box diagrams in terms of three-point functions.


\section{\label{sec::sqcd}Supersymmetric QCD}

\begin{figure}[t]
\begin{tabular}{cccc}
  \includegraphics[scale=1.1]{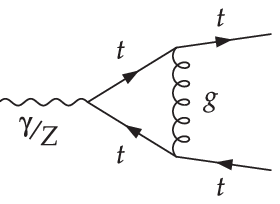}
  &
  \includegraphics[scale=1.1]{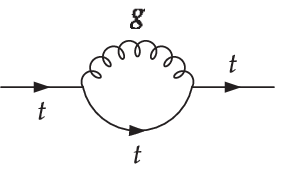}
  &
  \includegraphics[scale=1.1]{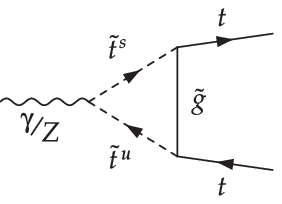}
  &
  \includegraphics[scale=1.1]{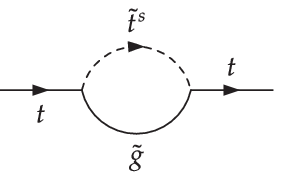}
  \\
  (a)
  &
  (b)
  &
  (c)
  &
  (d)
\end{tabular}
\caption{QCD and SQCD diagrams.
(a): Gluon contribution at the $t\overline{t}\gamma/Z$-vertex. 
(b): Gluon contribution to the top-quark selfenergy. 
(c): Gluino contribution at the $t\overline{t}\gamma/Z$-vertex. (d): Gluino
contribution to the top quark selfenergy.} 
\label{fig::diagQCD}
\end{figure}
In this Section we consider the effects from Supersymmetric QCD (SQCD) to 
the top quark threshold production. There are only four contributing Feynman
diagrams which are shown in Fig.~\ref{fig::diagQCD}.
The one-loop QCD corrections are known since long~\cite{Ka:1955} and the
corresponding matching coefficient is defined via the relation
\begin{eqnarray}
  J_V^i&=&\,\,\left(1+c_v^{(1)}\right)j_V^i
  \,\,=\,\,\left(1-2C_F\frac{\alpha_s}{\pi}\right)j_V^i\,.
  \label{eq::vectorcurrentmatching}
\end{eqnarray}
This effect can be incorporated in the helicity amplitude by a simple rescaling of the tree-level 
contributions
\begin{eqnarray}
  h_{I,V}^{\rm QCD} &=& h_{I,V}^{\gamma, \rm tree} a_{g}^\gamma+ h_{I,V}^{Z, \rm tree} a_{g}^Z\,.
  \label{eq::hIQCD}
\end{eqnarray}
Using the explicit expressions for $h_{I,V}^{\rm tree}$, one can already see that
the contribution of $a^Z_X$ to the relative correction $\Delta^X$ of the cross section
is in general suppressed by factor $0.08$ compared to the one
resulting from contribution of $a^{\gamma}_X$. 
For the QCD the coefficients $a^{\gamma/Z}_g$ read
\begin{eqnarray}
  a^{\gamma/Z}_g &=& c_v^{(1)}\,.
  \label{eq::cv}
\end{eqnarray}
Since we work in a supersymmetric framework we repeated the calculation of
$c_v^{(1)}$ within dimensional reduction~\cite{Siegel:1979wq}.
Although both the one-loop vertex corrections and the result for the wave
function counterterm are different from their counter parts in dimensional
regularization we observe that the result given in
Eq.~(\ref{eq::vectorcurrentmatching}) does not change. This is expected 
since at tree-level the strong coupling constant is absent.

The SQCD corrections can also be cast in the form of 
Eq.~(\ref{eq::hIQCD})

\begin{eqnarray}
  h_{I,V}^{\rm SQCD} &=& h_{I,V}^{\gamma, \rm tree} a_{\tilde g}^\gamma+
  h_{I,V}^{Z, \rm tree} a_{\tilde g}^Z\,, 
\end{eqnarray}
with
\begin{eqnarray}
  a_{\tilde g}^\gamma &=& \Gamma^\gamma_{V, \tilde g} + \delta Z^t_{V,\tilde g}\,,
  \nonumber\\
  a_{\tilde g}^Z &=& \Gamma^Z_{V, \tilde g} + \delta Z^t_{V,\tilde g}
  - \frac{3}{8s_w^2-3} \delta Z^t_{A,\tilde g} \,,
  \label{eq::a_gluino}
\end{eqnarray}
where $\Gamma^{\gamma/Z}_{V, \tilde g}$ represents the gluino
contribution to the vector part of the one-loop vertex normalized by the
corresponding tree level coupling. The wave function renormalization
constant $Z^{t}_{V/A,\tilde g}=1+\delta Z^{t}_{V/A,\tilde g}$ is defined in
the on-shell scheme and renders the $\gamma t\overline{t}$ and $Z
t\overline{t}$ vertex finite. The definition of both counter terms can be
found in Appendix \ref{sec::appnedixA}, where they are expressed in terms of
vector- ($V$) and axial-vector part ($A$) of the top quark selfenergy. The
subscript $\tilde g$ reminds that only the diagrams involving a gluino are
considered in each expression. 
Since the results are quite compact we present the analytical formulae for the
individual contributions of the right-hand side of
Eq.~(\ref{eq::a_gluino}). The contributions to the wave function
counterterm reads
\begin{align}
 \delta Z^t_{V,\tilde{g}}=&\overset{2}{\underset{s=1}{\sum}}\frac{\alpha_s}{6
   \pi
   m_t^2}\bigg\{-2m_t^2\bigg[-2m_{\tilde{g}}m_t\left(\Omega_{s\,1\,s\,2}+\Omega_{s\,2\,s\,1}\right)\nonumber\\ 
&+\left(m_t^2+m_{\tilde{g}}^2-m_{\tilde{t}_{s}}^2\right)
\left(\Omega_{s\,1\,s\,1}+\Omega_{s\,2\,s\,2}\right)\bigg]
B^\prime_0\left(m_t^2,m_{\tilde{g}}^2,m_{\tilde{t}_{s}}^2\right)\nonumber\\ 
&+\left(\Omega_{s\,1\,s\,1}+
  \Omega_{s\,2\,s\,2}\right)\Big[A_0\left(m_{\tilde{t}_{s}}^2\right)-A_0\left(m_{\tilde{g}}^2\right)\Big]\nonumber\\  
&+\left(m_{\tilde{g}}^2-m_t^2-m_{\tilde{t}_{s}}^2\right)
\Big(\Omega_{s\,1\,s\,1}+\Omega_{s\,2\,s\,2}\Big)B_0\left(m_t^2,m_{\tilde{g}}^2,m_{\tilde{t}_{s}}^2\right)\bigg\}\,, 
\nonumber\\
 \delta Z^t_{A,\tilde{g}}=&\overset{2}{\underset{s=1}{\sum}}\frac{\alpha _s
 }{6 \pi  m_t^2}\left(\Omega_{s\,1\,s\,1}-\Omega_{s\,2\,s\,2}\right) 
   \bigg\{A_0\left(m_{\tilde{g}}^2\right)-A_0\left(m_{\tilde{t}_{s}}^2\right)\nonumber\\
&-\left(m_t^2+m_{\tilde{g}}^2-m_{\tilde{t}_{s}}^2\right) 
   B_0\left(m_t^2,m_{\tilde{g}}^2,m_{\tilde{t}_{s}}^2\right)\bigg\}\,,
\label{EQ:DeltaZAGluino}\displaybreak[1]
\end{align}
and the vertex corrections are given by
\begin{align}
 \Gamma^{\gamma}_{V,\tilde{g}}=&\overset{2}{\underset{s=1}{\sum}}\frac{\alpha_s \big(\Omega_{s\,1\,s\,1}+\Omega_{s\,2\,s\,2}\big) }{9 \pi  m_t^2 \big(m_t^2+m_{\tilde{g}}^2-m_{\tilde{t}_{s}}^2\big)}\bigg\{2 m_t^2\left(m_t^2-m_{\tilde{t}_{s}}^2\right) B_0\left(4m_t^2,m_{\tilde{t}_{s}}^2,m_{\tilde{t}_{s}}^2\right)\nonumber\\ 
&+\tfrac{1}{2}\left(m_t^2+m_{\tilde{g}}^2-m_{\tilde{t}_{s}}^2\right)\Big[A_0\left(m_{\tilde{g}}^2\right)-A_0(m_{\tilde{t}_{s}}^2)+2m_t^2\Big]\nonumber\\ 
&-\tfrac{1}{2}\left[m_{\tilde{t}_{s}}^4-2 \left(m_t^2+m_{\tilde{g}}^2\right) m_{\tilde{t}_{s}}^2+\left(m_{\tilde{g}}^2-m_t^2\right)^2\right]B_0\left(m_t^2,m_{\tilde{g}}^2,m_{\tilde{t}_{s}}^2\right)\bigg\}\,,\nonumber\\
\Gamma^{Z}_{V,\tilde{g}}=&\overset{2}{\underset{s,\,u=1}{\sum}}\frac{\alpha_s \left[4s_w^2\left(\Omega_{s\,1\,u\,1}+ \Omega_{s\,2\,u\,2}\right)-3\Omega_{s\,1\,u\,1}\right] \left(\Omega_{u\,1\,s\,1}+\Omega_{u\,2\,s\,2}\right)}{9 \pi  m_t^2 \left(m_{\tilde{t}_{s}}^2+m_{\tilde{t}_{u}}^2-2 m_{\tilde{g}}^2-2 m_t^2\right) \left(8 s_w^2-3\right)}\nonumber\\
&\hspace{-1.2cm}\times \Bigg[\bigg\{ +\Big[ m_{\tilde{t}_{s}}^4-2 \left(m_t^2+m_{\tilde{g}}^2\right) m_{\tilde{t}_{s}}^2+\left(m_{\tilde{g}}^2-m_t^2\right)^2\Big]B_0\left(m_t^2,m_{\tilde{g}}^2,m_{\tilde{t}_{s}}^2\right)\nonumber\\
&\hspace{-0.1cm}+\Big[m_{\tilde{t}_{s}}^2+m_{\tilde{t}_{u}}^2-2\left(m_t^2+m_{\tilde{g}}^2\right)\Big]\nonumber\\
&\hspace{0.3cm}\times\tfrac{1}{2}\Big[A_0\left(m_{\tilde{g}}^2\right)-A_0\left(m_{\tilde{t}_{s}}^2\right)+2m_t^2\left(1+B_0\left(4m_t^2,m_{\tilde{t}_{s}}^2,m_{\tilde{t}_{u}}^2\right)\right)\Big]\nonumber\\
&\hspace{-0.1cm}+\Big[m_t^2\left(m_{\tilde{t}_{s}}^2+m_{\tilde{t}_{u}}^2+2\left(m_{\tilde{g}}^2-m_t^2\right)\right)-\tfrac{1}{4}\left(m_{\tilde{t}_{s}}^2-m_{\tilde{t}_{u}}^2\right)^2\Big]\nonumber\\
&\hspace{0.3cm}\times B_0\left(m_t^2,m_{\tilde{t}_{s}}^2,\tfrac{1}{2}m_{\tilde{t}_{s}}^2+\tfrac{1}{2}m_{\tilde{t}_{u}}^2-m_t^2\right)\bigg\}+\bigg\{s\leftrightarrow u\bigg\}\Bigg]\,.
\label{EQ:aZGluinoUR}
\end{align}

In Eqs.~(\ref{EQ:DeltaZAGluino}) and~(\ref{EQ:aZGluinoUR}) we
introduced the abbreviations
$\Omega_{ijkl} = U_{ij} U^\star_{kl}$ where $U_{ij}$ are the elements of the
top squark mixing matrix (cf. Appendix~\ref{sec::appnedixB}).
The conventions for the functions $A_0$ and $B_0$ are adapted from
Ref.~\cite{Passarino:1978jh,Hahn:1998yk} where explicit results can be
found. Further $B_0^\prime$ is the defined as derivative of $B_0$ with respect
to the first argument. 
Our analytic formulae are in agreement with Ref.~\cite{Su:2001pi} where the result
has been expressed in terms of a one-dimensional integral assuming a real
mixing matrix for the top squarks. 

It is instructive to consider the limit where all SUSY particles have
a common mass $m_{\rm SUSY}$. In this limit the above formulae are
simplified significantly. In particular, the result becomes independent of the
matrix elements $U_{ij}$ and $\delta Z_{A,\tilde{g}}^t=0$. We furthermore have
$a_{\tilde{g}} = a^{\gamma}_{\tilde{g}} = a^{Z}_{\tilde{g}}$ which reads
\begin{eqnarray}
  a_{\tilde{g}}(m^2_{\textnormal{SUSY}})&=&\frac{4\alpha_s}{9m_t^2\pi}\bigg\{\tfrac{1}{2}m_t^2-\tfrac{3}{2}m_t^4B_0^\prime(m_t^2,m^2_{\textnormal{SUSY}},m^2_{\textnormal{SUSY}})\nonumber\\
  &&+\left(m^2_{\textnormal{SUSY}}-m_t^2\right)\left[B_0(m_t^2,m^2_{\textnormal{SUSY}},m^2_{\textnormal{SUSY}})-B_0(4m_t^2,m^2_{\textnormal{SUSY}},m^2_{\textnormal{SUSY}})\right]\bigg\}
  \nonumber\\
&=&\frac{\alpha_s}{45\pi}\bigg\{y^2+\tfrac{16}{21}y^3+\tfrac{1}{2}y^4+\tfrac{76}{231}y^5+\mathcal{O}(y^6)\bigg\}\,.
\label{eq::aGluinoMSUSYLimit}
\end{eqnarray}
After the second equal sign we have expanded the result in terms of
$y=m_t^2/m_{\rm SUSY}^2$.

Let us in the following discuss the numerical effects of the one-loop QCD
and SQCD corrections. For $m_t=173.1$~GeV and $\alpha_s^{(6)}(m_t)=0.108$
the QCD corrections amount to $\Delta^{\rm QCD}=18.3\%$ (corresponding to
$\alpha_s^{(5)}(m_Z)=0.1176$) and thus constitute the largest contribution. 

\begin{figure}[t]
  \centering
  \hspace*{-1cm}
  \includegraphics[scale=0.70]{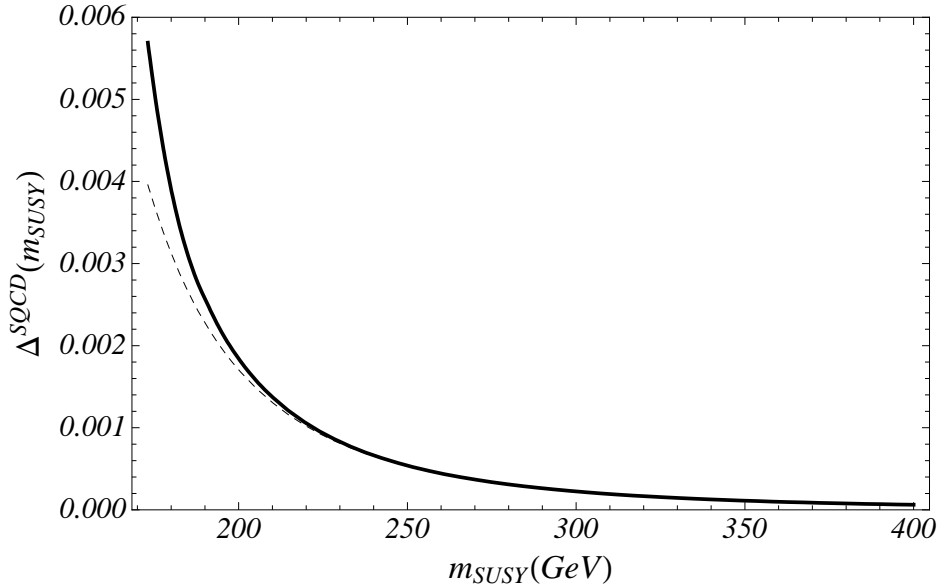}
  \caption{\label{fig::aGluinoMSUSY}
    $\Delta^{\rm SQCD}=2a_{\tilde g}$ computed from
    Eq.~(\ref{eq::aGluinoMSUSYLimit}) as a function of $m_{\rm SUSY}$.
    The solid line represents the exact result and the dashed curve
    the expansion including terms up to order $(m_t^2/m_{\rm SUSY}^2)^5$.
  }
\end{figure}

\begin{figure}[t]
\centering
\hspace*{-1cm}
\begin{tabular}{cc}
  \includegraphics[scale=0.40]{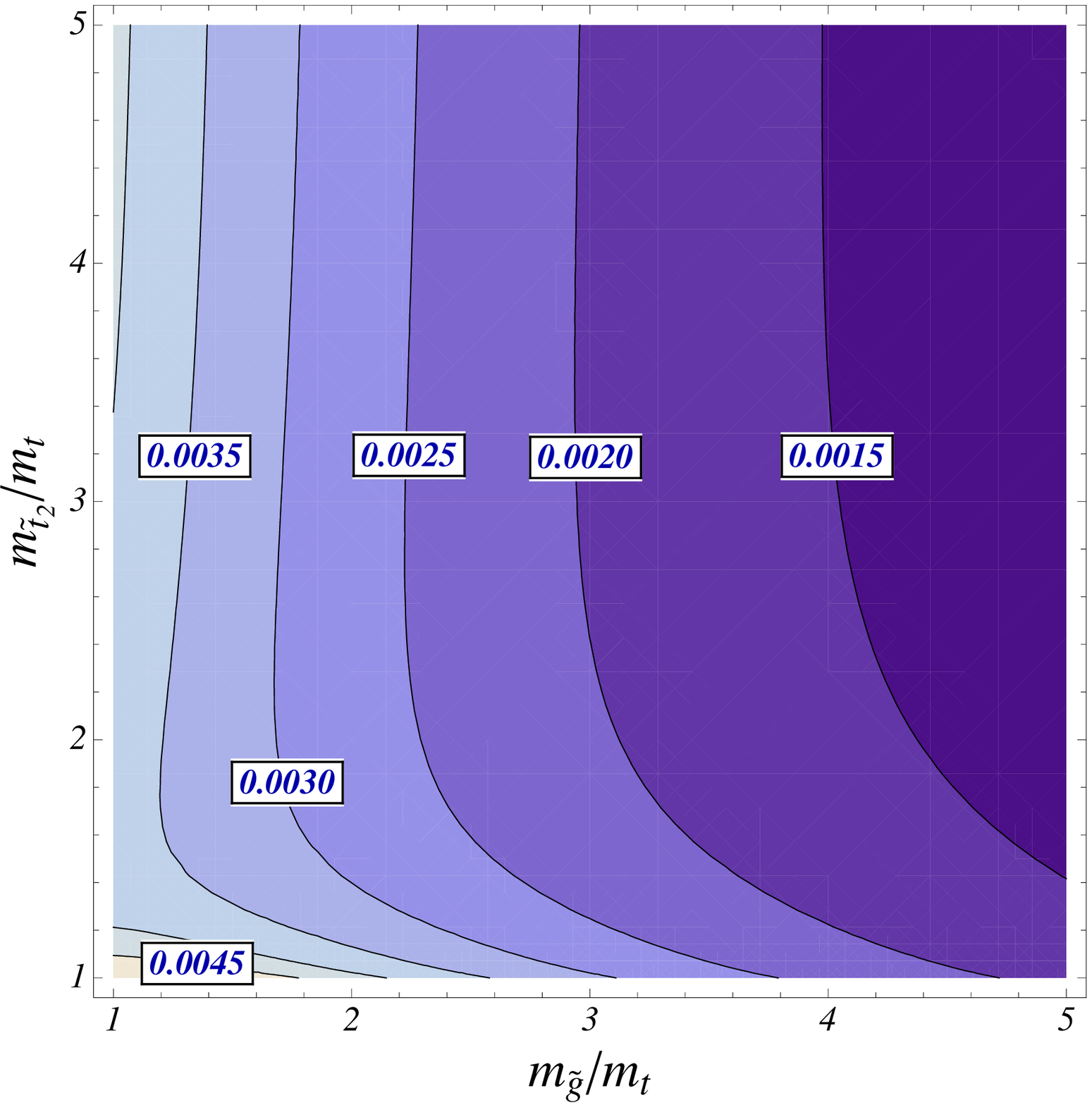}
  &
  \includegraphics[scale=0.40]{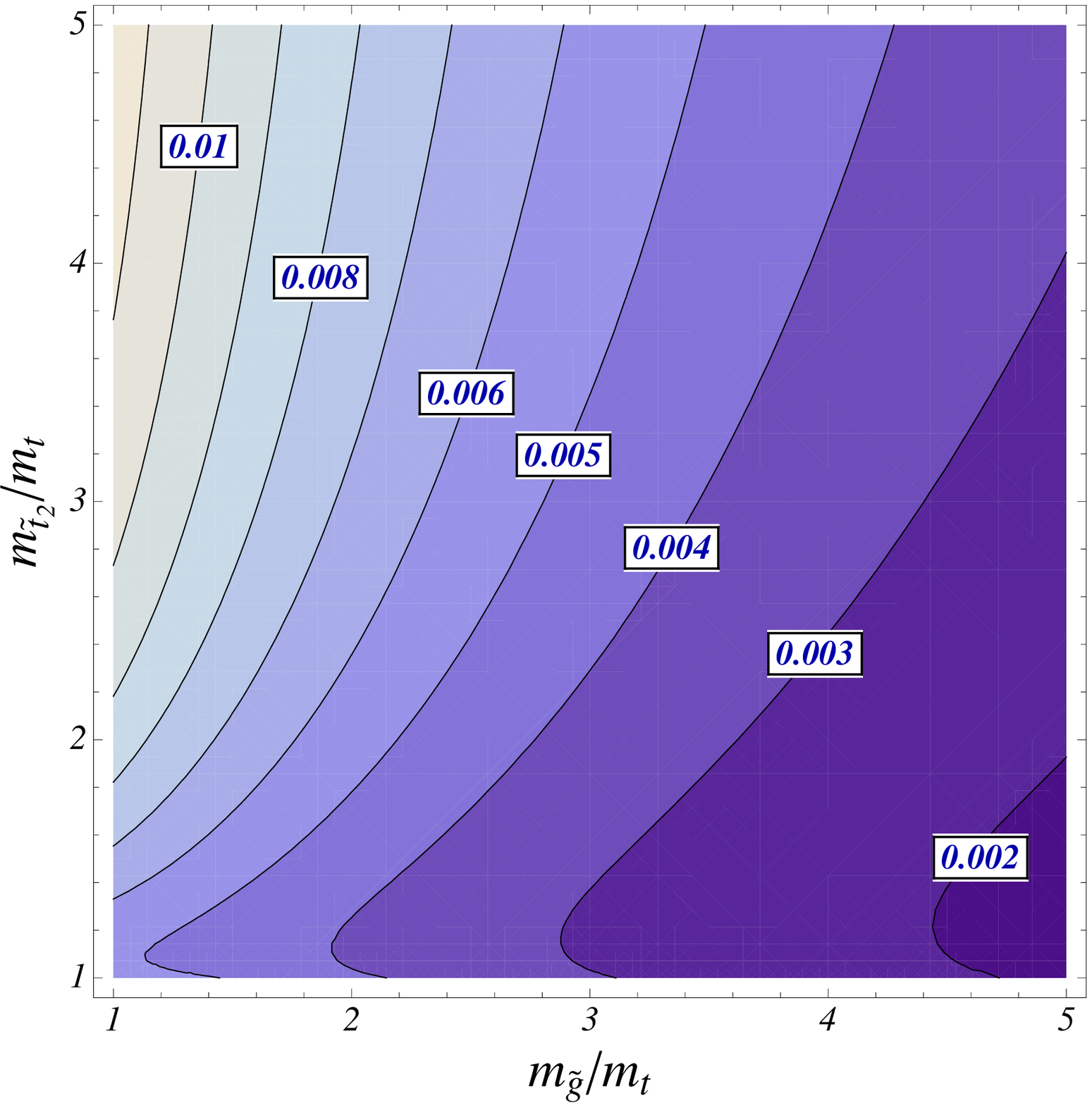}
  \\
  (a) $\Delta^{\textnormal{SQCD}}(\theta_{\tilde{t}}=0)$
  &
  (b) $\Delta^{\textnormal{SQCD}}(\theta_{\tilde{t}}=\pi/4)$
  \\
  \includegraphics[scale=0.40]{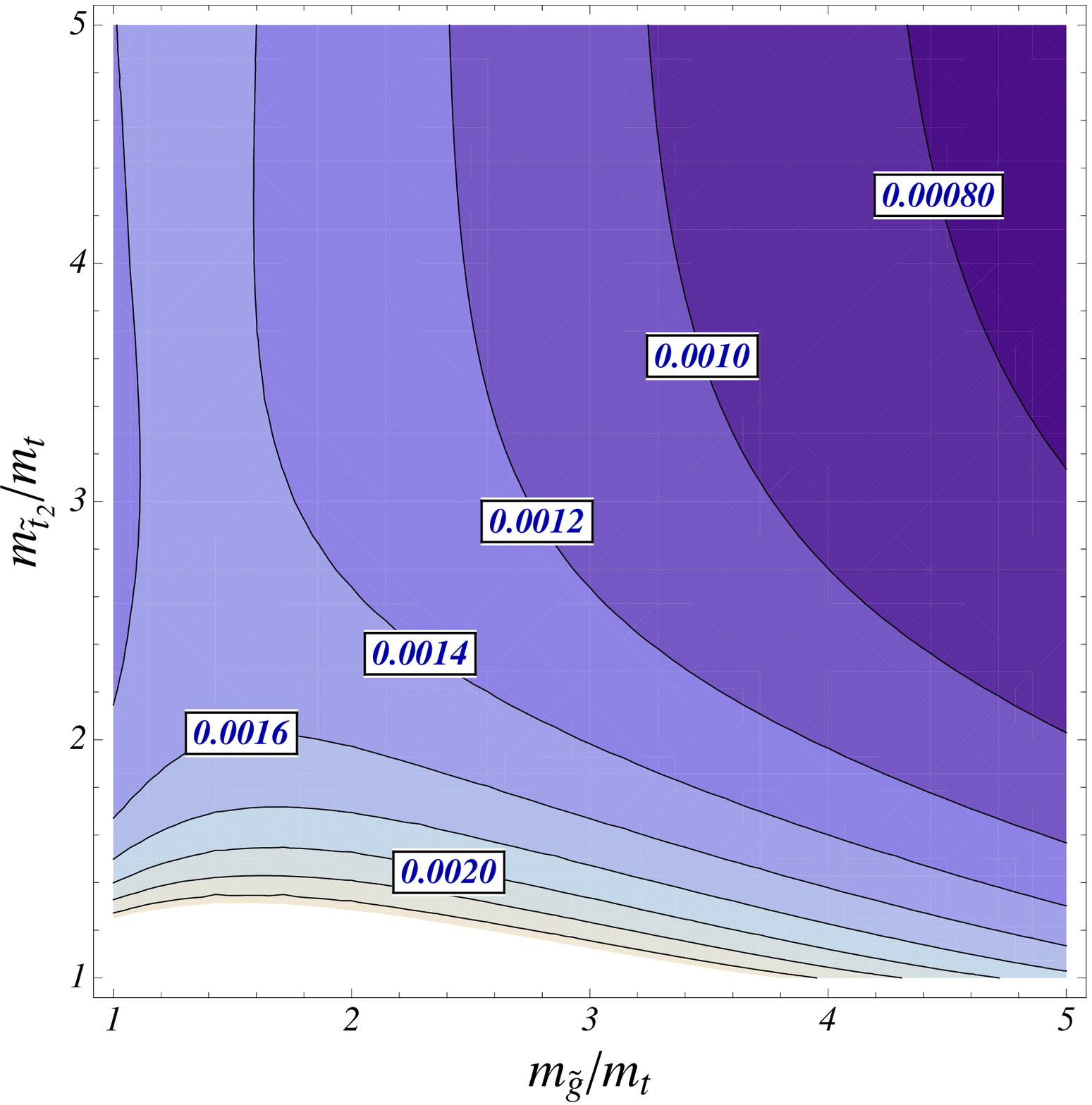}
  &
  \includegraphics[scale=0.40]{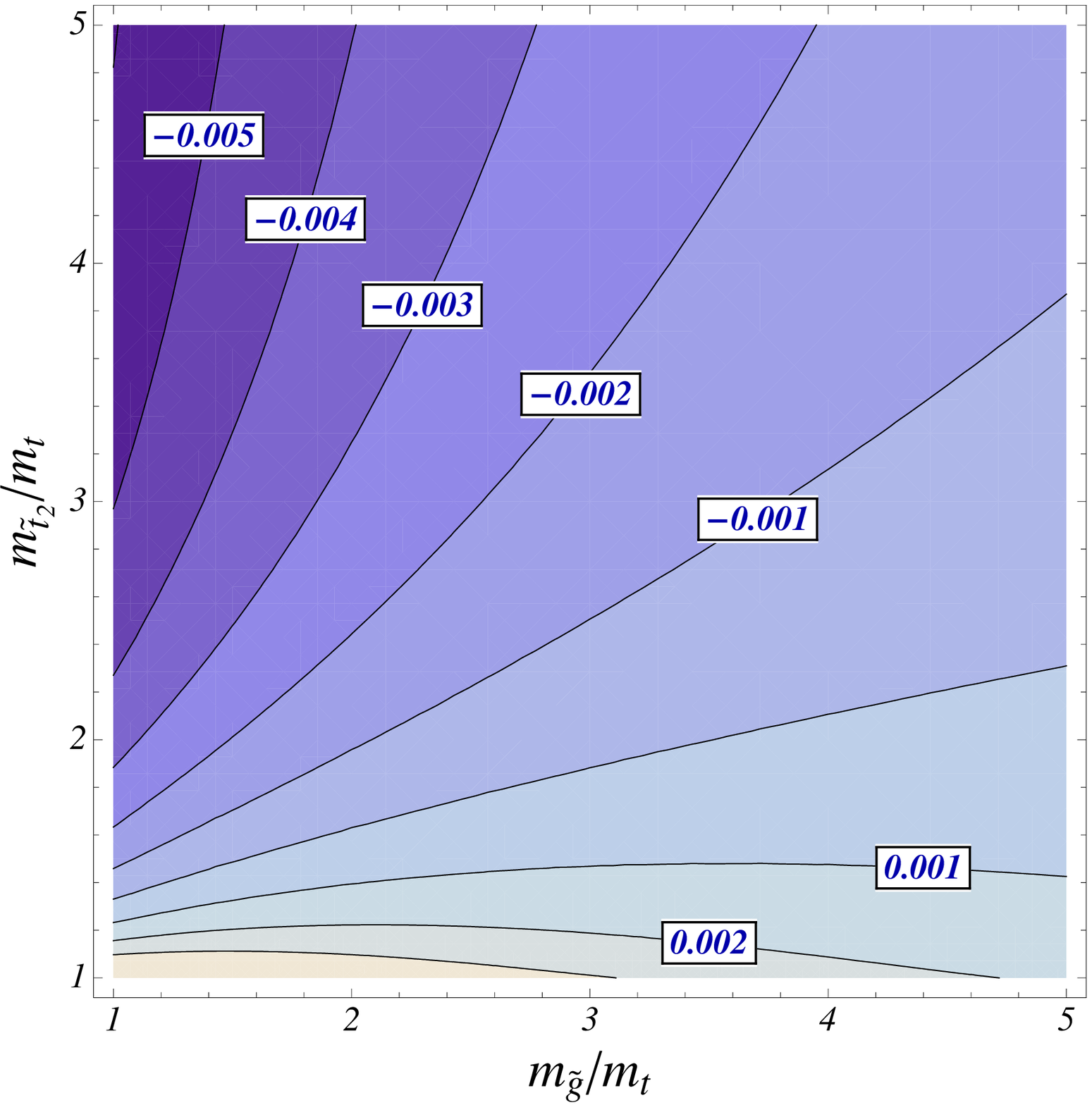}
  \\
  (c) $\Delta^{\textnormal{SQCD}}(\theta_{\tilde{t}}=\pi/2)$
  &
  (d) $\Delta^{\textnormal{SQCD}}(\theta_{\tilde{t}}=3\pi/4)$
\end{tabular}
\caption{$\Delta^{\textnormal{SQCD}}$ as a function of $m_{\tilde{t}_2}$ and $m_{\tilde{g}}$ (normalized to the
  top quark mass) for $m_{\tilde{t}_1}=m_{t}$ and different values
  of the mixing angle $\theta_{\tilde{t}}$.}
\label{fig::DeltaSUSYQCD}
\end{figure}

In the simplified scenario described by Eq.~(\ref{eq::aGluinoMSUSYLimit}) one
obtains the SQCD corrections as shown in Fig.~\ref{fig::aGluinoMSUSY}. 
From the figure one can see that for $m_{\rm SUSY}>200\rm GeV$ the expansion
agrees well with the exact result showing a relative deviation below $10$\%. 
For all $m_{\rm SUSY}>m_t$ the relative correction to the threshold cross section stays below 0.6\%.
The size of the SQCD corrections in a non-universal SUSY mass scenario is
shown in Fig.~\ref{fig::DeltaSUSYQCD} where $\Delta^{\rm SQCD}$ is plotted as
a function of $m_{\tilde{t}_2}$ and $m_{\tilde{g}}$ for $m_{\tilde{t}_1}=m_t$.  
Since our results are $\pi$-periodic in $\theta_{\tilde{t}}$ we have chosen
for illustration the four values $\theta_{\tilde{t}} \in
\{0,\tfrac{\pi}{4},\tfrac{\pi}{2},\tfrac{3\pi}{4}\}$. 
The figures show, that only for light masses of the second top squark
($m_{\tilde{t}_2}\lesssim 2m_t$)  $\Delta^{\rm SQCD}$ can have a strong
dependence on $m_{\tilde{t}_2}$. 

In general one observes corrections below 1\% which become negligible for
large masses of the SUSY particles. A correction factor above 1\% is only
observed for $\theta_{\tilde{t}}=\pi/4$ and relatively light gluino masses
of the order of the top quark mass which are excluded within the
MSSM~\cite{Amsler:2008zzb}.


\section{\label{sec::ew}Electroweak corrections in the THDM and the MSSM}

QCD corrections only affect the $\gamma t\overline{t}/Z t\overline{t}$
vertex. On the other hand, electroweak corrections require also the 
inclusion of the $ e^+ e^-\gamma/e^+ e^-Z$ vertex and furthermore of gauge
boson self 
energy and box contributions which are necessary in order to arrive at a 
finite and gauge parameter independent result.
Typical Feynman diagrams contributing to the individual building blocks
are shown in Fig.~\ref{fig::diagSM} for the SM and in Fig.~\ref{fig::diagMSSM}
for the MSSM.
Due to the renormalization procedure (we follow Ref.~\cite{Bohm:1986rj,Hollik:1993cg})
also $W$ boson and fermion selfenergy
contributions have to be computed which are also shown in
Figs.~\ref{fig::diagSM} and~\ref{fig::diagMSSM}. They are used in order to
render the four building blocks individually finite which is quite convenient to deal with.

\newcommand{\picturescale}{0.9}

\begin{figure}[t]
\centering
\begin{tabular}{cccc}
  \includegraphics[scale=\picturescale]{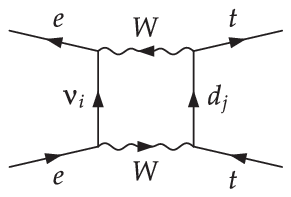}
  &
  \includegraphics[scale=\picturescale]{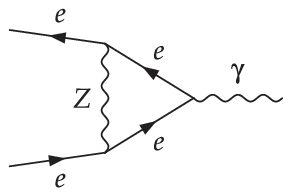}
  &
  \includegraphics[scale=\picturescale]{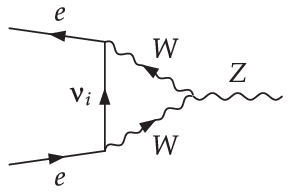}
  &
  \includegraphics[scale=\picturescale]{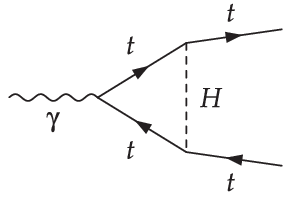}
  \\
  (a)
  &
  (b)
  &
  (c)
  &
  (d)
  \\
  \includegraphics[scale=\picturescale]{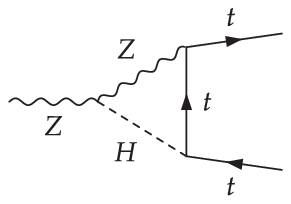}
  &
  \includegraphics[scale=\picturescale]{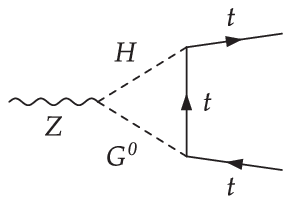}
  &
  \includegraphics[scale=\picturescale]{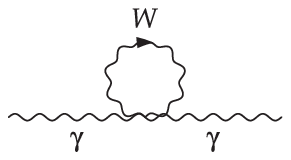}
  &
  \includegraphics[scale=\picturescale]{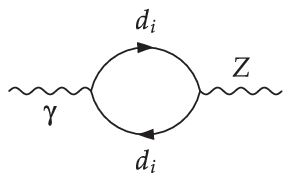}
  \\
  (e)
  &
  (f)
  &
  (g)
  &
  (h)
  \\
  \includegraphics[scale=\picturescale]{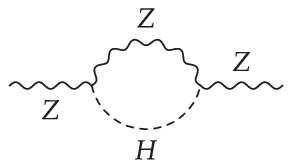}
  &
  \includegraphics[scale=\picturescale]{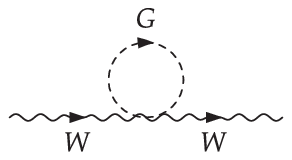}
  &
  \includegraphics[scale=\picturescale]{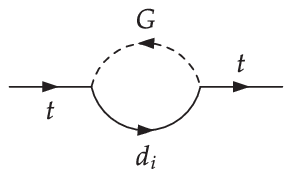}
  &
  \includegraphics[scale=\picturescale]{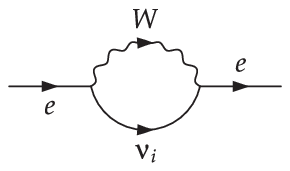}
  \\
  (i)
  &
  (j)
  &
  (k)
  &
  (l)
  \\
\end{tabular}
\caption{Typical Feynman diagrams contributing to $\Delta^{\rm SM}$.}
\label{fig::diagSM}
\end{figure}

\begin{figure}[t]
\centering
\begin{tabular}{cccc}
  \includegraphics[scale=\picturescale]{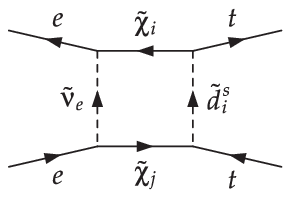}
  &
  \includegraphics[scale=\picturescale]{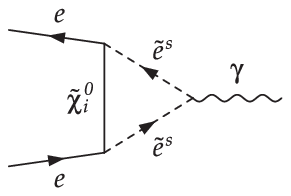}
  &
  \includegraphics[scale=\picturescale]{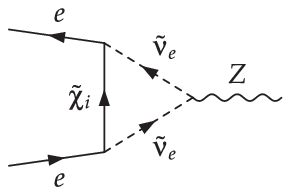}
  &
  \includegraphics[scale=\picturescale]{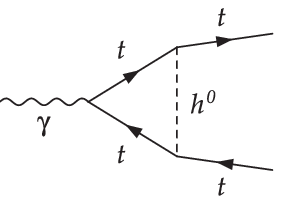}
  \\
  (a)
  &
  (b)
  &
  (c)
  &
  (d)
  \\
  \includegraphics[scale=\picturescale]{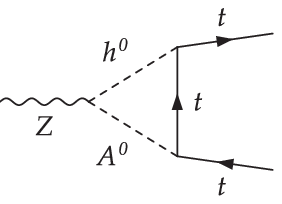}
  &
  \includegraphics[scale=\picturescale]{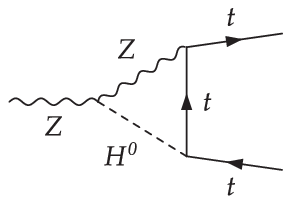}
  &
  \includegraphics[scale=\picturescale]{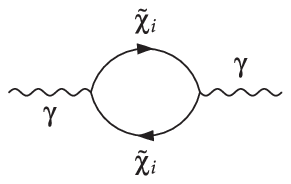}
  &
  \includegraphics[scale=\picturescale]{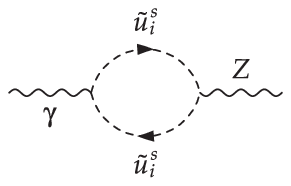}
  \\
  (e)
  &
  (f)
  &
  (g)
  &
  (h)
  \\
  \includegraphics[scale=\picturescale]{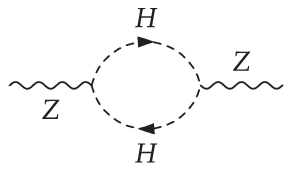}
  &
  \includegraphics[scale=\picturescale]{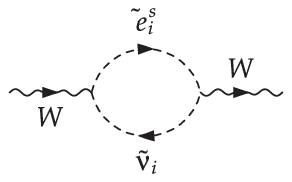}
  &
  \includegraphics[scale=\picturescale]{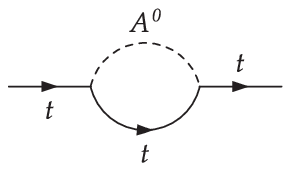}
  &
  \includegraphics[scale=\picturescale]{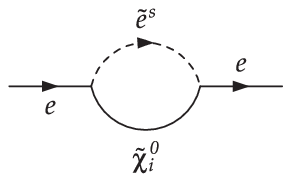}
  \\
  (i)
  &
  (j)
  &
  (k)
  &
  (l)
  \\
\end{tabular}
\caption{Typical Feynman diagrams contributing to $\Delta^{\rm MSSM}$.}
\label{fig::diagMSSM}
\end{figure}

In a first step we have used our set-up in order to compute the SM
contribution. We find complete agreement with
Refs.~\cite{Grzadkowski:1986pm,Guth:1991ab,Hoang:2006pd}.
Afterwards the THDM model has been considered
and the results from Ref.~\cite{Denner:1991tb} have been reproduced.\footnote{In
Ref.~\cite{Denner:1991tb} the expression for $a^Z_{Zh^0}$ is proportional to
$\cos(\beta-\alpha)$ which should be replaced by $\sin(\beta-\alpha)$.}
Let us note that for vanishing bottom quark mass the corrections in the THDM model can easily be
obtained from the analytical results for Higgs- and Goldstone boson contribution calculated in the SM by adjusting the
coupling factors and boson masses in the loop diagrams.

Results for the MSSM are not yet available in the literature. However,
it is possible to compare our results for the vector boson selfenergies with
Ref.~\cite{Hollik:1998md} where the top quark production has been
considered above the threshold.
As far as the box contribution is concerned, new kind of diagrams occur in the MSSM where the
electron and positron in the initial state are not part of the same fermion
line (and similarly for the top quarks in the final state),
cf. Fig.~\ref{fig::diagMSSM}(a).
Due to the different tensor structure, originating from the Majorana character
of charginos and neutralinos, it is not straightforward to process these
contributions with our set-up.  
On the other hand, it is possible to extract the relative correction to the
cross section at the threshold by taking the limit 
$s\to 4m_t^2$ since these diagrams only involve heavy particles 
inside the loop. However, due to the numerical  properties of the loop functions \cite{Hahn:1998yk}
the limit can not be taken naively. Instead we evaluate the
result of Ref.~\cite{Hollik:1998md} for the box contribution above threshold
and extrapolate to $s=4m_t^2$. In this way we obtain the threshold
contribution for the new box 
diagrams with three significant digits which is sufficient for the phenomenological analysis.  
We have applied the same procedure for the SM box contributions which provides
both a cross check on our analytical calculation and the very procedure for
extracting the threshold contribution.   

Due to the occurrence of many different masses and mixing angles the remaining
general expression is quite lengthy in the case of the MSSM.   
Thus, in the following we will only discuss the numerical effects.   
In Ref. \cite{NZ:2009pck} a package  is provided which allows the numerical
evaluation of the corrections described in this paper.   
It uses {\tt Mathematica} as front-end and calls Fortran for the
time-consuming parts of the calculation.   
In addition an interface to SPheno \cite{Porod:2003um} is provided, which
generates numerical values for the masses and mixing angles on the basis of a
certain SUSY breaking scenario.   

In the numerical discussion we will restrict ourselves to the SUSY
breaking scenario based on minimal supergravity (mSUGRA) and use the
Snowmass Points and Slopes (SPS)~\cite{Allanach:2002nj,AguilarSaavedra:2005pw}
in order get an impression 
of size of the corrections.
In addition to the five mSUGRA parameters $m_0$, $m_{1/2}$, $\tan\beta$,
$A_0$ and $\text{sgn}(\mu)$ (cf. Tab.1) which serve as input for the
spectrum generator we use the following input values for the remaining
SM parameters
\cite{:2009ec,Abbaneo:2009,Kuhn:1998ze}\footnote{Following
  Ref. \cite{Denner:1991kt} we replace light fermion contributions to the
  derivative of the photon vacuum polarization function by
  $\Delta\alpha^{(5)}_{\text{had}}(m_Z)$ and
  $\Delta\alpha_{\text{lep}}(m_Z)$.}  
\begin{align}
  m_W&=80.40\,\text{GeV},&m_Z&=91.1876\,\text{GeV},     &c_w^2&=m_W^2/m_Z^2,\nonumber\\
  m_{t}&=173.1\,\text{GeV}, & m_{b}&=4.2\,\text{GeV}, &\alpha^{-1}&=137.036,\nonumber\\
  \Delta\alpha^{(5)}_{\text{had}}(m_Z)&=277.45\times10^{-4}, & \Delta\alpha_{\text{lep}}(m_Z)&=314.97\times10^{-4}.
\label{eq::SMparameters}
\end{align}

In a first step our {\tt Mathematica} program transfers the input values to
the spectrum generator {\tt SPheno}~\cite{Porod:2003um} 
which produces numerical values for all unknown MSSM parameters relevant for our analysis.
The output is automatically imported into {\tt Mathematica} and afterwards
used in order to evaluate the THDM or MSSM corrections. 
More details about the functionality of our package is provided via the usual
{\tt Mathematica} internal documentation and example files which in addition
automatically generate the plots and tables shown in this paper. 

\begin{table}[t]
\centering
\scalebox{1}[1]{
\begin{tabular}{c|c|c|c|c|c|c|}\cline{2-7}
&\multicolumn{4}{c}{Points}&\multicolumn{2}{|c|}{Slopes}\\
\hline
\hline
\multicolumn{1}{|c||}{Label}& $m_0$ & $m_{1/2}$ & $A_0$ & $\tan{\beta}$& $m_0$ & $A_0$\\
\hline
\hline
\multicolumn{1}{|c||}{SPS1a'} & 70    & 250       & $-$300  &  10  & - &  - \\
\hline
\multicolumn{1}{|c||}{SPS1a}  & 100   & 250       & $-$100  &  10  & $0,4m_{1/2}$ &  $-0,4m_{1/2}$ \\
\hline
\multicolumn{1}{|c||}{SPS1b}  & 200   & 400       &    0  &  30  & - &  - \\
\hline
\multicolumn{1}{|c||}{SPS2}   & 1450  & 300       &    0  &  10  & $2m_{1/2}+850$ & 0 \\
\hline
\multicolumn{1}{|c||}{SPS3}   & 90    & 400       &    0  &  10  & $0,25m_{1/2}-10$ & 0 \\
\hline
\multicolumn{1}{|c||}{SPS4}   & 400   & 300       &    0  &  50  & - &  - \\
\hline
\multicolumn{1}{|c||}{SPS5}   & 150   & 300       & $-$1000 &  5   & - &  - \\
\hline
\end{tabular}
}
\caption{Input values for the SPS scenarios as defined in references \cite{Allanach:2002nj,AguilarSaavedra:2005pw}. All masses are given in GeV and $\textnormal{sgn}(\mu)=1$. }
\label{tab::SPSDefinitions}
\end{table}

The numerical impact of the corrections in different mSUGRA scenarios can be seen
in Tab.~\ref{tab::SPS} where $\Delta^{\rm SM\,EW}$, $\Delta^{\rm THDM\,EW}$
and $\Delta^{\rm MSSM\,EW}$ are evaluated for several SPS points\footnote{We
  add EW to the superscript in order to make clear that only electroweak and
  no strong corrections are considered}.   
Note that $\Delta^{\rm SM\,EW}$ varies since the SM Higgs boson is identified
with the lightest MSSM Higgs boson.   

\begin{table}
\scalebox{0.9}[0.9]{
\setlength{\doublerulesep}{0.5mm}
\setlength{\tabcolsep}{1mm}
\renewcommand{\arraystretch}{1.1}
\centering
\hspace*{1cm}
\begin{tabular}{c|R[.][.]{1}{3}|R[.][.]{1}{3}|R[.][.]{1}{3}|R[.][.]{1}{3}|R[.][.]{1}{3}|R[.][.]{1}{3}|R[.][.]{1}{3}|}\cline{2-8}
 \null  & \multicolumn{1}{m{1,3cm}}{SPS1a} & \multicolumn{1}{|m{1,3cm}}{SPS1a'} & \multicolumn{1}{|m{1,3cm}}{SPS1b} & \multicolumn{1}{|m{1,3cm}}{SPS2} & \multicolumn{1}{|m{1,3cm}}{SPS3} & \multicolumn{1}{|m{1,3cm}}{SPS4} & \multicolumn{1}{|m{1,3cm}|}{SPS5}\\
 \hline
 \hline
\multicolumn{1}{|c||}{$\Delta ^{\text{SM EW}}$} & 0.15199574430263782 & 0.1508928851239338 & 0.14869143858026393 & 0.14856436630005368 & 0.1493739221066284 & 0.15037658750621538 & 0.14928989397503967\\
\hline
\multicolumn{1}{|c||}{$\Delta ^{\text{THDM EW}}$} & 0.09739603552016611 & 0.09649447811742255 & 0.09340634924465764 & 0.0906870961241125 & 0.0931357206179212 & 0.09927101883995555 & 0.09399906162905693\\
\hline
\multicolumn{1}{|c||}{$\Delta ^{\text{MSSM EW}}$} & 0.09647116224717922 & 0.09568010799134066 & 0.09343098677403545 & 0.08918521715965788 & 0.09265602741247984 & 0.10057006361858124 & 0.09356396284995777\\
\hline
\end{tabular}

\renewcommand{\arraystretch}{1}
}
\caption{Numerical values for $\Delta ^{\text{X EW}}$ $X \in \{{\rm SM, THDM, MSSM}\}$ for variouse SPS scenarios.}
\label{tab::SPS}
\end{table}

\begin{figure}[t]
\begin{tabular}{c}
\includegraphics[scale=0.85]{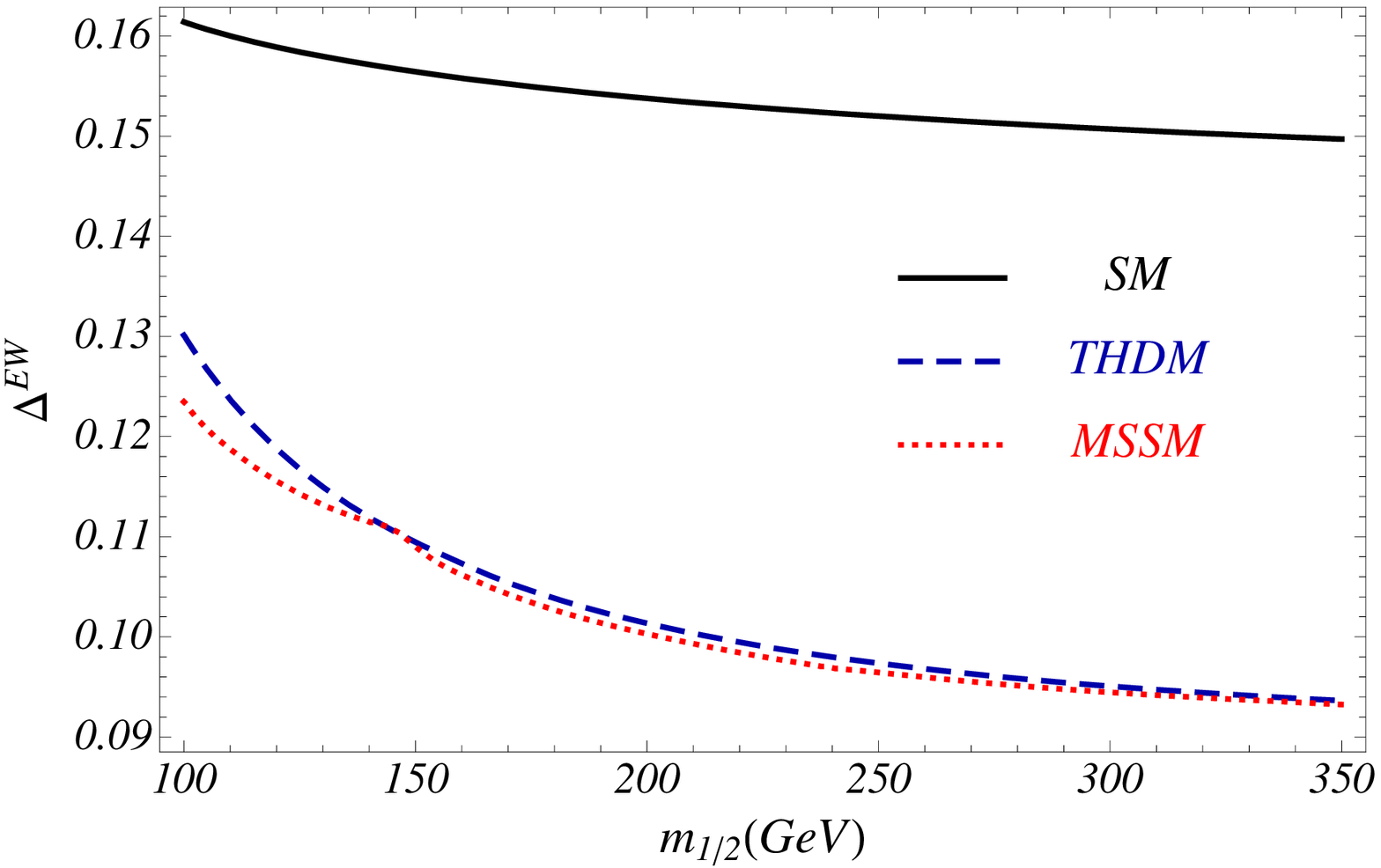}
\\
(a)
\\
\includegraphics[scale=0.85]{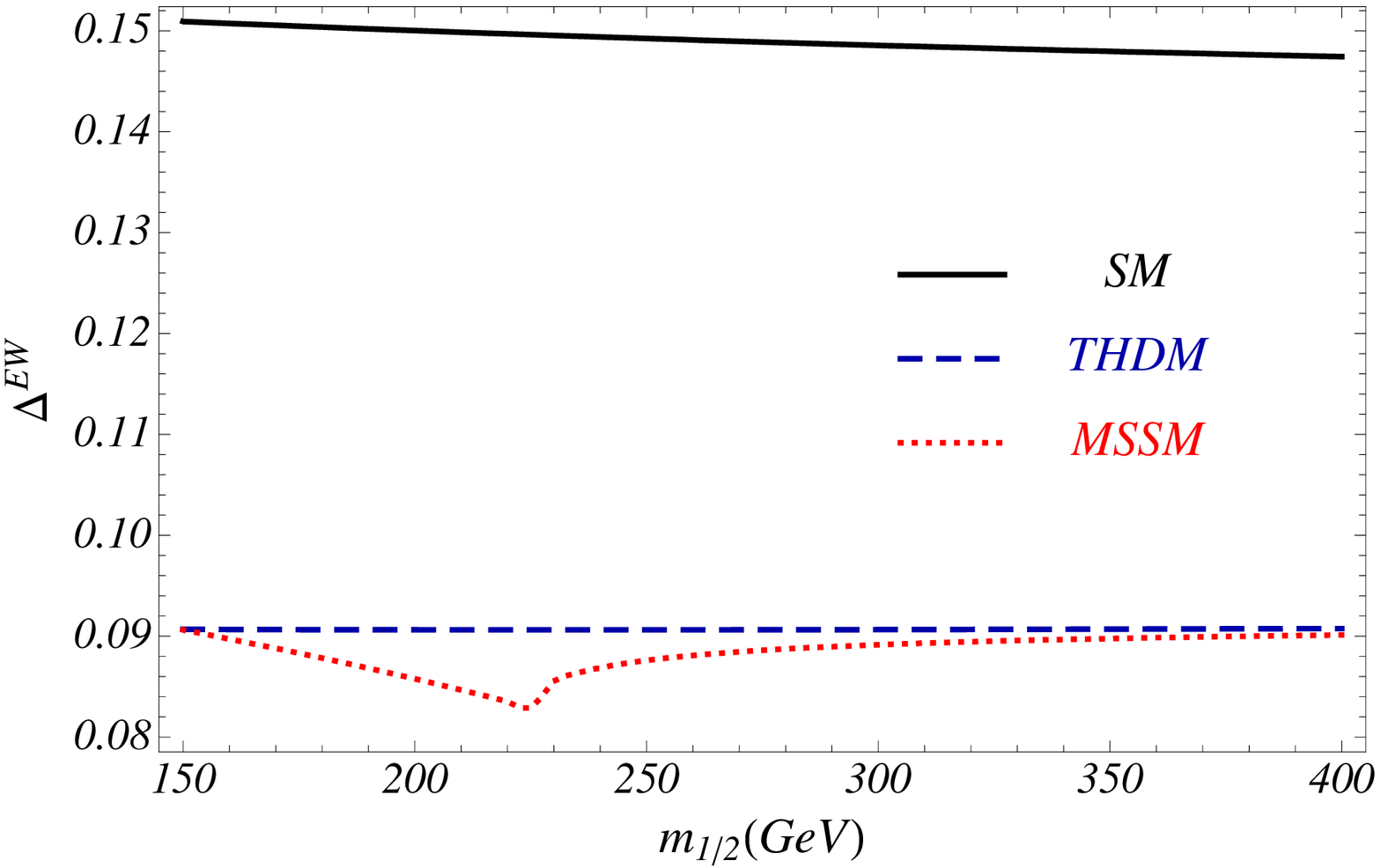}
\\
(b)
\end{tabular}
\caption{$\Delta^{\textnormal{X EW}}$ for $X \in \{{\rm SM, THDM, MSSM}\}$ as function of the unified mSUGRA gaugino mass $m_{1/2}$ for (a) SPS1a and (b) SPS2.}
\label{figs::SPS12}
\end{figure}

\begin{figure}[t]
\hspace*{-0.2cm}
\begin{tabular}{cc}
\includegraphics[scale=0.42]{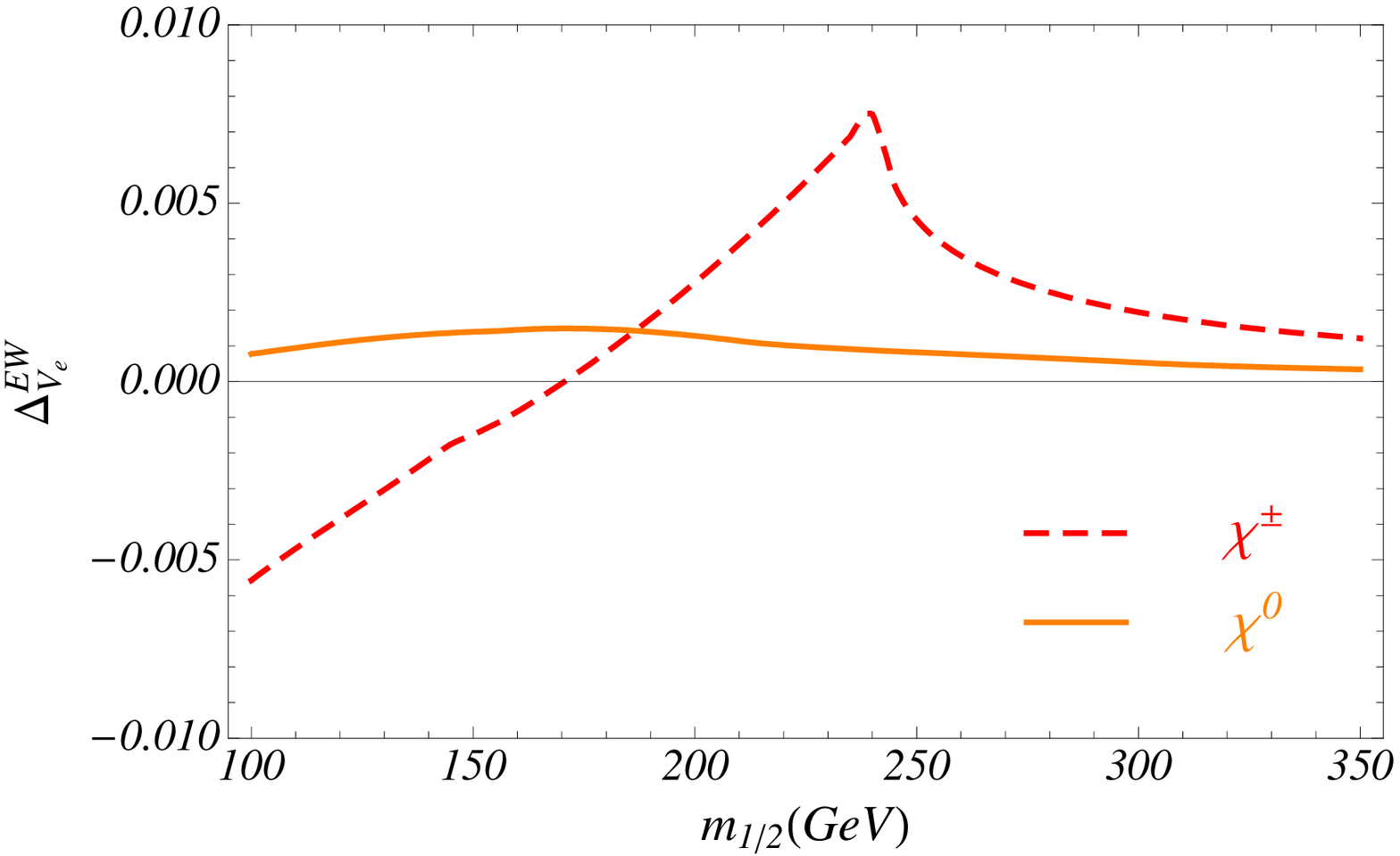}
&
\includegraphics[scale=0.42]{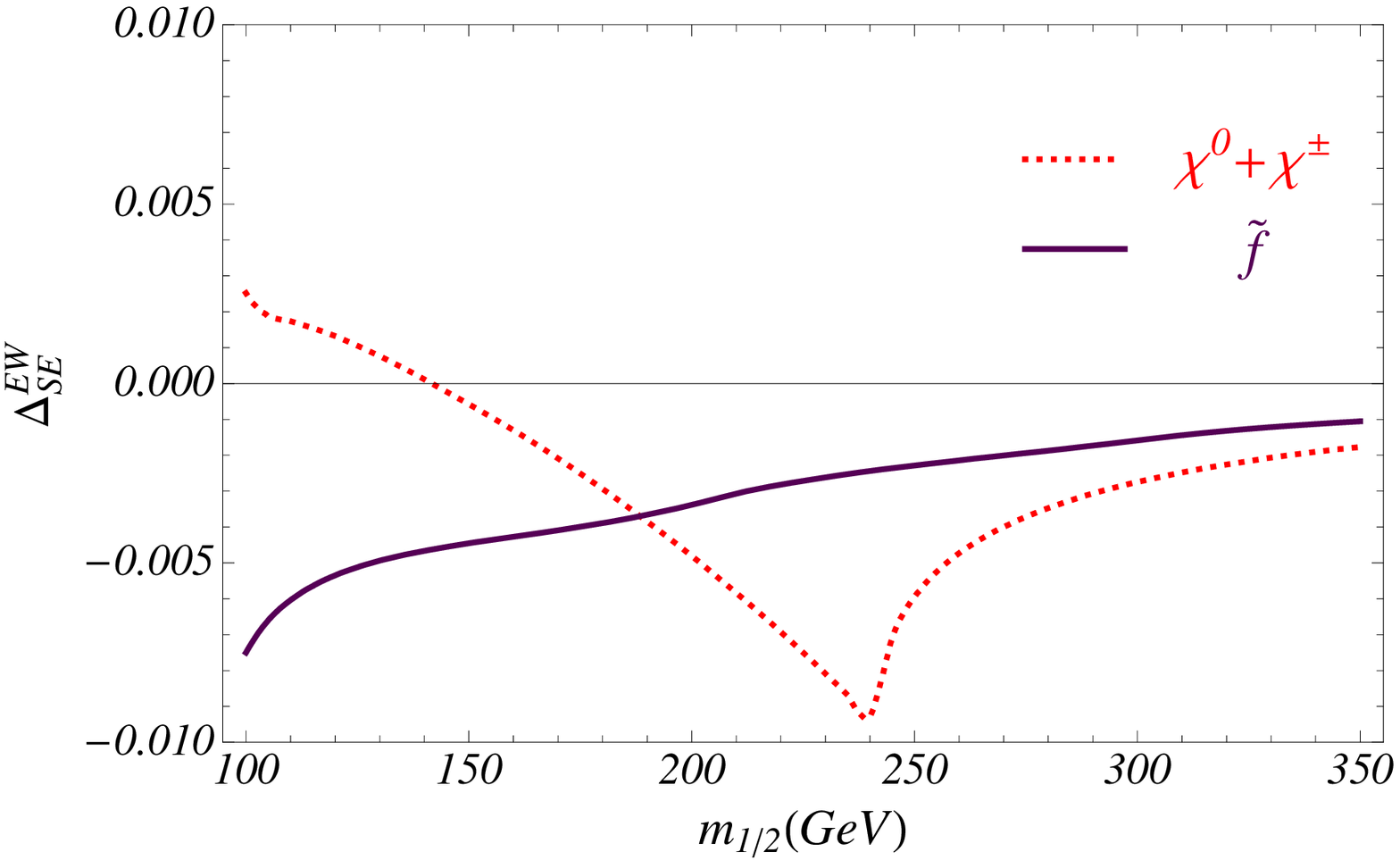}
\\
(a)
&
(b)
\\
\includegraphics[scale=0.42]{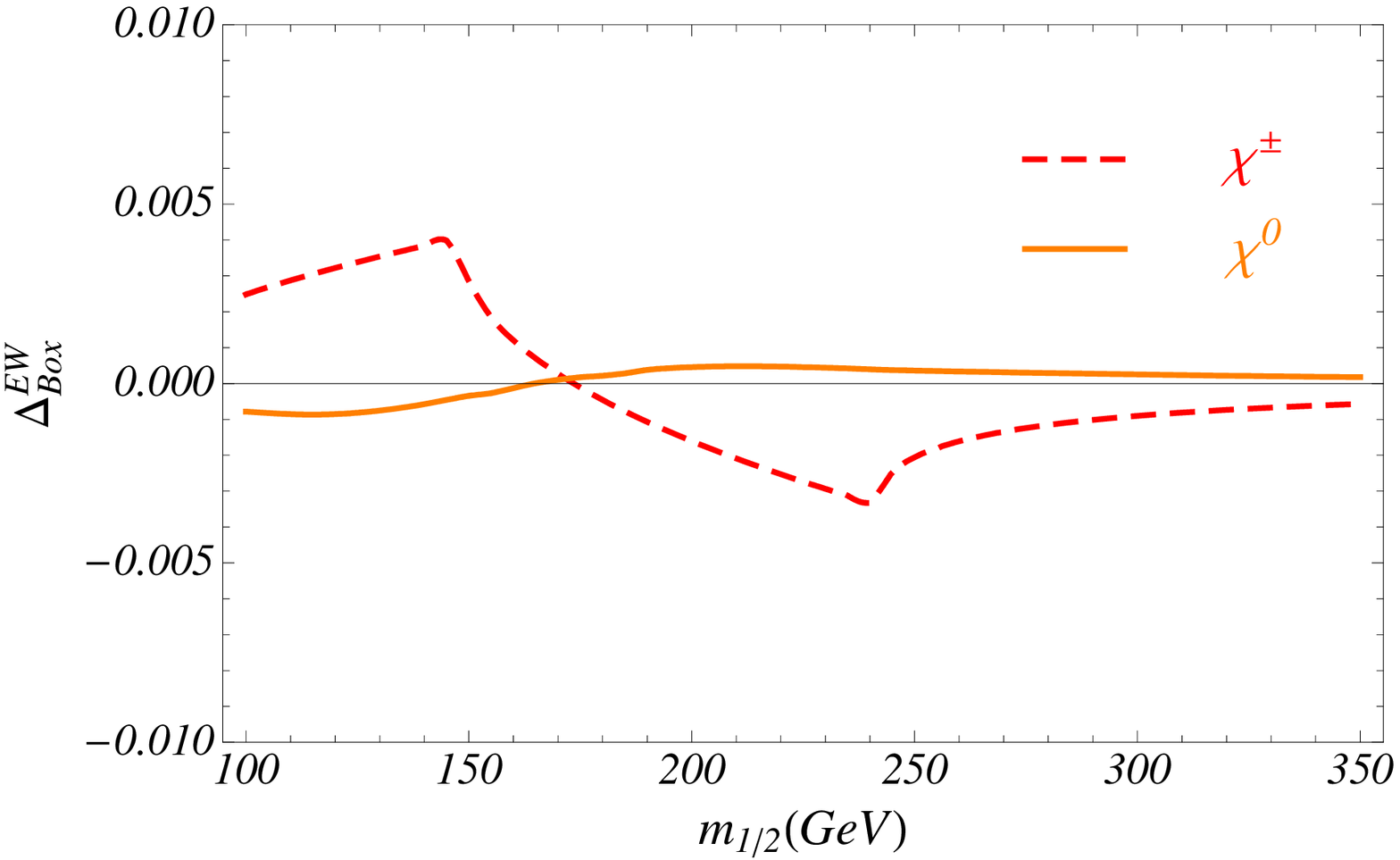}
&
\includegraphics[scale=0.42]{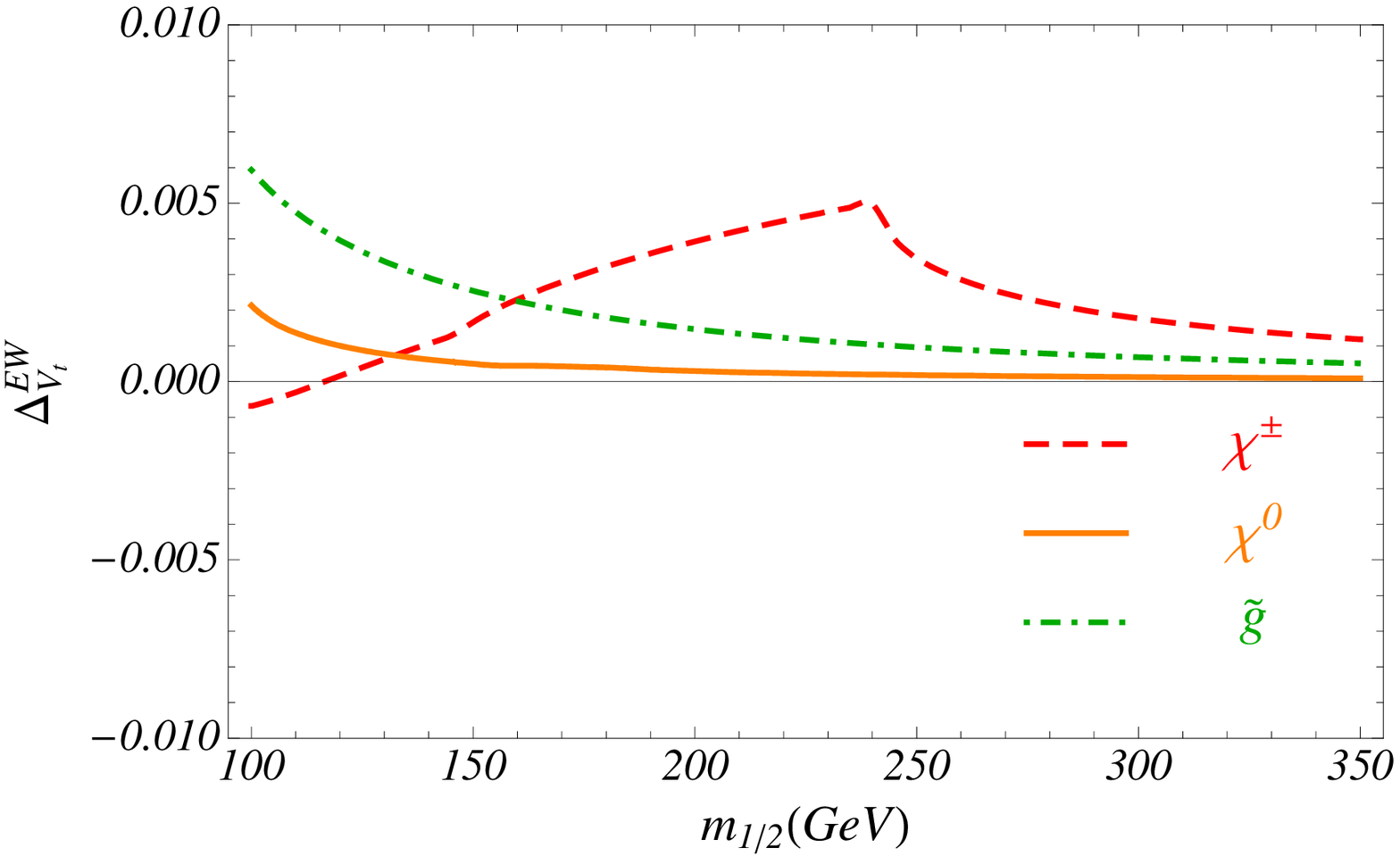}
\\
(c)
&
(d)
\\
\end{tabular}
\caption{Contributions of the building blocks to $\Delta^{\textnormal{MSSM EW}}$ 
in the MSSM as function of the unified gaugino mass $m_{1/2}$ for SPS1a: (a) electron vertex,  (b) vector boson selfenergie, (c) box and (d) top quark vertex. In (d) the SQCD corrections are shown for comparison.}
\label{fig::SPS1Details}
\end{figure}

The SM corrections amount to a sizeable shift of about 15\% which get
reduced by roughly 5\% to 6\% in the case of the THDM. 
The main reason for this reduction is the smaller coupling of the top quark to the light Higgs boson.
At the same time only numerically small contributions arise from the diagrams
involving heavy Higgs bosons.
In Tab.~\ref{tab::SPS} one observes only a marginal difference between
the THDM and the MSSM. It is thus instructive to have a closer look at
the depedence on $m_{1/2}$ as suggested by the SPS scenarios. For
illustration we show in Fig.~\ref{figs::SPS12} the comparison of
$\Delta^{\text{SM EW}}$, $\Delta^{\text{THDM EW}}$ and $\Delta^{\text{MSSM EW}}$
for SPS1 and SPS2. In both cases we observe only small corrections beyond the 
THDM, i.e. from the neutralino and chargino sector of the MSSM. Larger deviations of the order of $0.5$\% 
are only observered for those values of $m_{1/2}$ where the
corresponding chargino masses are close to the top quark mass.
This becomes clear in Fig. 7 where we show the correction of the finite building blocks separately for the case of SPS1a.
One can see a relatively strong variation in the dashed curve which shows the contributions from the charginos.
It is interesting to note that the peak around $m_{1/2}\approx150\text{GeV}$ in $\Delta_{\text{Box}}$ is clearly
visible in Fig. \ref{figs::SPS12} (a) whereas a cancelation among the various parts occures for the peak close to $m_{1/2}\approx 250\text{GeV}$.
For comparison we plot in Fig.\ref{fig::SPS1Details}(d) the contribution from SQCD (dashed-dotted).
For $m_{1/2}\gtrsim 150\text{GeV}$ it is smaller than the chargino contribution. 
Corrections above $0.5$\% are only reached for relatively small values of $m_{1/2}$ which corresponds to small values of the gluino mass.\\
Let us finally mention that we performed our calculation in the THDM and MSSM for finite bottom quark mass and investigated possible large corrections for higher values of $\tan{\beta}$.
However, even for the mSUGRA scenario SPS4 where $\tan{\beta}=50$, 
the result for massless bottom quark is $\Delta^{\textnormal{MSSM EW}}_{m_b=0}=0.099$, thus
the effect of finite bottom quark mass  adds 0.002 to $\Delta^{\textnormal{MSSM EW}}_{m_b=0}$
(see Tab. \ref{tab::SPS}).

\section{\label{sec::conclusions}Conclusions}
In this paper we investigated the complete weak and strong one-loop
corrections within the MSSM to the threshold production of top-quark pairs at
a future $e^+e^-$ linear collider. 
For the SM, THDM, QCD and SQCD corrections we confirmed the results in the
literature, the genuine supersymmetric electroweak corrections are new. 

As far as the numerical importance is concerned, the electroweak SM
corrections amount uo to $+15$\% for light Higgs masses. 
After extending the Higgs sector we observe for the SPS scenarios a screening
of about $-5$\% to $-6$\% in the THDM (type II). 
The pure supersymmetric corrections from the chargino, neutralino and the
strong sector are below $1$\% in most of the parameter space.  


\bigskip
\noindent
{\large\bf Acknowledgements}\\
We would like to thank Christian Schappacher for sharing his program
and knowledge about {\tt FormCalc}, Thomas Hahn for his support,
and J.H.~K\"uhn for valuable discussions.
This work was supported by the DFG through SFB/TR~9 and BMBF O5HT4VK4VKAI3.

\appendix
\section{\label{sec::appnedixA}{On-shell counterterms}}
In this appendix we discuss the definition of the counterterms appearing in
Eq.~(\ref{eq::a_gluino}).
In order to define the fermionic on-shell counterterms one needs the
coefficient functions of the tensor decomposition from the corresponding
fermion selfenergy: 
\begin{equation}
\Sigma(q,m)=m\,\Sigma_s(q^2,m)+\fmslash{q}\,\Sigma_v(q^2,m)+\fmslash{q}\gamma^5\,\Sigma_a(q^2,m).
\label{eq::SigmaTensorDecomposition}
\end{equation}
They can be extracted with the help of the following projections:
\begin{subequations}
\label{eq::ProjectionsOnTensorCoefficients}
\begin{align}
\Sigma_s(q^2,m)&=\frac{1}{4m}tr\left\{\Sigma(q,m)\right\},\\
\Sigma_v(q^2,m)&=\frac{1}{4q^2}tr\left\{\fmslash{q}\,\Sigma(q,m)\right\},\\
\Sigma_a(q^2,m)&=\frac{1}{4q^2}tr\left\{\gamma^5\fmslash{q}\,\Sigma(q,m)\right\}.
\end{align}
\end{subequations}
The wave function counterterms are then given by
\begin{eqnarray}
\label{eq::FermionicCounterTerms}
\delta Z^f_V&=&-\Sigma_v(m_f^2,m_f)-2m_f^2\left.\frac{\partial }{\partial
    q^2}\bigg[\Sigma_v(q^2,m_f)+\Sigma_s(q^2,m_f)\bigg]\right|_{q^2=m_f^2}\,,\\ 
\delta Z^f_A&=&\Sigma_a(m_f^2,m_f)\,.
\end{eqnarray}

\section{\label{sec::appnedixB}{Mixing matrices}}
Let us for definiteness provide in this Appendix the definition of the mixing
matrix $\mathbf{U}_{\tilde{f}}=(U_{ij})$ used in Eq. \ref{EQ:DeltaZAGluino}
and \ref{EQ:aZGluinoUR}. 
We work with flavor diagonal sfermion mixings, where the left and right handed
sfermion fields $\tilde{f}_L$ and $\tilde{f}_R$ are connected to the mass
eigenstates $\tilde{f}_1$ and $\tilde{f}_2$ via 
\begin{eqnarray}
\label{eq::SfermionMixingMatrix}
\left(\begin{array}{c}\tilde{f}_1\\\tilde{f}_2\end{array}\right)=\mathbf{U}_{\tilde{f}}\left(\begin{array}{c}\tilde{f}_L\\\tilde{f}_R\end{array}\right)\,. 
\end{eqnarray}
The $2\times2$ mixing matrix $\mathbf{U}_{\tilde{f}}$ diagonalizes the mass
matrix of the corresponding sfermion $\tilde{f}$: 
\begin{eqnarray}
\label{eq::DiagonalisationOfSfermionMassMatrix}
\mathbf{U}_{\tilde{f}}\,\mathbf{m}^2_{\tilde{f}}\,\mathbf{U}^{\dagger}_{\tilde{f}}=
\left(\begin{array}{cc}m^2_{\tilde{f}_1}&0\\0&m^2_{\tilde{f}_2}\end{array}\right)\,.  
\end{eqnarray}
In the case where the mass matrix $\mathbf{m}^2_{\tilde{f}}$ contains only
real entries, one can choose  $\mathbf{U}_{\tilde{f}}$ to be orthogonal. For
its parameterization only one angle $\theta_{\tilde{f}}$ is needed and the
transformation from mass to gauge eigenstates can be written as follows: 
\begin{eqnarray}
\label{eq::RealSfermionMixing}
\tilde{f}_1&=&\tilde{f}_L\cos{\theta_{\tilde{f}}}+\tilde{f}_R\sin{\theta_{\tilde{f}}}\,,\nonumber\\
\tilde{f}_2&=&\tilde{f}_R\cos{\theta_{\tilde{f}}}-\tilde{f}_R\sin{\theta_{\tilde{f}}}\,.
\end{eqnarray}



\end{document}